%% file: main.tex
\documentclass[reprint,amsmath,amssymb,amsbsy,prb,footinbib]{revtex4-2}
\usepackage{graphicx,color}
\usepackage{braket}
\usepackage{bm}
\usepackage{color}
\usepackage{ulem}
\usepackage{verbatim}
\usepackage[colorlinks=true, citecolor=blue, linkcolor=blue]{hyperref}
\usepackage{times}

\newcommand{\mUpsilon}{{\mathcal O}}
\newcommand{\mathI}{{I}}

\input{mydef}

{\par\noindent{\em #1:\ }}%
{~\rule{2mm}{2mm}\par\bigskip}

\begin{document}
\title{Hidden chiral symmetry for kagome lattice and its analogs}

\author{Tomonari Mizoguchi}
\email{mizoguchi@rhodia.ph.tsukuba.ac.jp}
\affiliation{Department of Physics, University of Tsukuba, Tsukuba, Ibaraki 305-8571, Japan}

\author{Yasuhiro Hatsugai}
\email{hatsugai.yasuhiro.ge@u.tsukuba.ac.jp}
\affiliation{Department of Physics, University of Tsukuba, Tsukuba, Ibaraki 305-8571, Japan}

\date{\today}
\begin{abstract}
Chiral symmetry plays an essential 
role in condensed matter physics.
In tight-binding models,
it is often attributed to 
bipartite lattice structures,
and its typical consequence is the ``particle-hole symmetric" band structures, that is, the positive and negative eigenenergies appear in a pairwise manner. 
In this work, 
we address the chiral symmetry for
non-bipartite lattice models with flat bands.
Our argument relies on what we call the molecular-orbital representation, by which we can guarantee the existence of the flat band.
We show that the chiral symmetry is preserved for the non-flat bands when the molecular orbitals are normalized and divided into two sets within which they are non-overlapping.
The chiral operator is constructed by the same molecular orbitals as the Hamiltonian.
This results in the characteristic relation between the chiral operator and the Hamiltonian, which we call the Pythagoras relation. 
We present the examples of such models, i.e., the kagome lattice and its one-dimensional analog, and address the characteristics originating from the chiral symmetry, such as the emergence of the topological edge modes.
We also show an example where the numbers of molecular orbitals in each set are different from each other, resulting in multiple flat bands with different energies.
\end{abstract}

\maketitle
\section{Introduction}
Characterization of the band structures 
by symmetries 
is a fundamental aspect of condensed matter physics. 
Among various symmetries, chiral symmetry plays important roles in various condensed matter systems. 
It becomes important especially 
when considering simple tight-binding models.
In such systems, 
chiral symmetry is often attributed to 
bipartite lattice structures, 
thus is often called the sublattice symmetry as well.
In the presence of chiral symmetry, the band structure becomes ``particle-hole symmetric", namely, the positive and negative eigenenergies appear in a pairwise manner.
The chiral symmetry leads to various exotic phenomena, 
such as Dirac fermions~\cite{Nielsen1981,Nielsen1981_2,Hatsugai2010,Hatsugai2013,Koshino2014,Wakao2020}, 
flat bands~\cite{Sutherland1986,Lieb1989},
chiral zero modes in random systems~\cite{Brouwer2002},
and topological phases~\cite{Altland1997,Ryu2002,Schnyder2008,Kitaev2009,Ryu2010}.

In many of the previous works, 
the chiral symmetric models are realized either
by using the bipartite nature of the lattice or
by using some additional ingredients such as the strong spin-orbit coupling~\cite{Nakai2023}.
In this paper, 
we show the construction scheme of the tight-binding models 
on non-bipartite lattices 
that have exact flat bands
and preserve the hidden chiral symmetry.
Here we use what we call the molecular-orbital (MO) representation~\cite{Hatsugai2011,Hatsugai2015,Mizoguchi2019,Mizoguchi2020,Hatsugai2021,Mizoguchi2021_skagome,Mizoguchi2022,Kuroda2022,Mizoguchi2023}, which we have developed to describe generic flat band models.
We show that the chiral symmetry is preserved for the non-flat bands when the MOs satisfy two conditions, namely, (i)~the MOs are normalized and (ii)~they are divided into two sets within which they are non-overlapping.
The chiral symmetry originates from 
the chiral symmetry of the overlap matrix.
The chiral operator for these models 
has several distinct features compared with the conventional ones, as the chiral operator is also constructed by the MOs. 
As such, the chiral operator is closely related to the Hamiltonian itself, which results in the characteristic Pythagoras relation between 
the Hamiltonian and the chiral operator. 
As examples, we consider the kagome lattice and its one-dimensional analog (the modified saw-tooth lattice), and 
address their characteristics resulting from the chiral symmetry, such as the emergence of the topological edge modes.
Additionally, we also consider the square-ocgaton lattice,
 which serves as an example where the numbers of MOs 
 in each set are different from each other.
We show that there are mulitiple flat bands with different energies
due to the hidden chiral symmetry and the MO-number imbalance.

The rest of this paper is organized as follows.
In Sec.~\ref{sec:mo}, we review the
MO representation and its basic properties.
In Sec.~\ref{sec:chiral}, we present our main result, namely, we show that the MO model preserves the chiral symmetry when the MOs satisfy certain conditions.
We further elaborate the 
consequence of the hidden chiral symmetry,
namely the particle-hole-symmetric spectrum and the properties of the zero modes in 
Sec.~\ref{sec:p-h}.
In Sec.~\ref{sec:sawtooth} and Sec.~\ref{sec:kagome},
we present the concrete examples of the modified saw-tooth lattice and the kagome lattice, respectively.
We address the characteristic phenomena resulting from the chiral symmetry, such as the emergence of the topological edge modes. 
In Sec.~\ref{sec:sqo}, we consider the square-octagon lattice,
which is an example of the model with the MO-number imbalance of each set.
Finally, we summarize this paper in Sec.~\ref{sec:summary}.

\section{Molecular-orbital representation \label{sec:mo}}
We begin with reviewing 
the basic framework of the MO representation, 
which we have developed in the literatures~\cite{Hatsugai2011,Hatsugai2015,Mizoguchi2019,Mizoguchi2020,Hatsugai2021,Mizoguchi2021_skagome,Mizoguchi2022,Kuroda2022,Mizoguchi2023}. 
We consider the tight-binding model for the spinless single-orbital fermions.
We write the annihilation operator of 
the fermion at site $i$ ($i=1, \cdots, N$) 
as $c_i$
; we call $c_i$s' as the annihilation operator of the atomic orbitals (AOs).
The key idea of the MO representation is to define the 
creation operator of the MO by the
linear combination of that of the AOs: 
\begin{align}
  C^\dagger_I = \sum_{i=1}^N c^\dagger_i \psi_{i,I} 
\end{align}
with $I = 1, \cdots, M$.
Althoguh defined as such, 
the coefficient $\psi_{i,I}$ can be regarded as 
the wavefunction of the MO $I$.
See also Appendix~\ref{app1_1}.
Aligning $C^\dagger_I$ and $c^\dagger_i$ in the form of the row vector, $\bm{C}^\dagger = \left(C^\dagger_1, \cdots, C^\dagger_M \right)$ and $\bm{c}^\dagger = \left(c^\dagger_1, \cdots, c^\dagger_N \right)$,
we have
\begin{align}
\bm{C}^\dagger = \bm{c}^\dagger \Psi,
\end{align}
where 
\begin{align}
\Psi 
  &=
  (
\psi_1,
\cdots ,
  \psi_M) 
  \\
  &= 
  \left(
  \begin{array}{ccc}
    \psi_{1,1}&\cdots&\psi_{1,M} \\
    \vdots &\ddots&\vdots \\
    \psi_{N,1} &\cdots&\psi_{N,M}
  \end{array}
  \right) \in M(N,M)\nonumber, 
  \\
  \psi_I &= \mvecthree{\psi_{1,I}}{\vdots}{\psi_{N,I}}\in M(N,1),
  \ I=1,\cdots,M,
\end{align}
where  $M(a,b)$ is a set of $a \times b$ matrices and
$\bm{\psi}_I 
$ is the $N$-component 
column vector, which we identify to the $N\times 1$ matrix.

We consider the tight-binding Hamiltonian
which can be written by using $C_I$,
\begin{align}
  \mathcal{H}= \sum_{I,J=1}^{M} h_{IJ} C^\dagger_{I}C_J
= \bm{C}^\dagger h \bm{C}
= \bm{c}^{\dagger} 
  {H} \bm{c}, \label{eq:Ham_MO}
\end{align}
with $H:= \Psi h \Psi^\dagger$.
Here, $h$ is the $M\times M$ Hermitian matrix.
Although $h$ can be any in a generic framework, 
we focus on the case where $h$ is the identity matrix for the sake of arguing the chiral symmetry.
Thus, in this paper, we have 
$
{H}:= \Psi \Psi^\dagger$.
The single-particle eigenvalues and eigenstats for $\mathcal{H}$ can be obtained by solving the eigenvalue problem for 
the matrix $H$.
See also Appendix \ref{app1_1}.

The model constructed as such has two remarkable features. 
First, the existence 
of the degenerate 
zero-energy modes is guaranteed if $M < N$.
This is because $\Psi^\dagger\in M(M,N)$, 
hence there exist at least $Z:=N-M$ column vectors $\bm{\varphi}_{\ell}$ ($\ell = 1, \cdots, Z$) that satisfies 
$\Psi^\dagger \bm{\varphi}_{\ell} = 0$.
Such vectors are the zero-energy eigenvectors of 
$
{H}$ since $
{H}\bm{\varphi}_{\ell} = \Psi (\Psi^\dagger
\bm{\varphi}_{\ell}) = 0$.
Second, the non-zero energy eigenvalues can be obtained by solving the eigenvalue equation of the $M\times M $ matrix, $\mUpsilon= \Psi^\dagger \Psi$, which we refer to 
as the overlap matrix. 
Indeed, the matrix elements of $\mUpsilon$ correspond 
to the overlap between the MOs, namely, $[\mUpsilon]_{IJ} = \psi_I^\dagger \psi_J$. 
The relation between $
{H}$ and $\mUpsilon$ is as follows. Let 
$\bm{u}_{\ell}$ be a normalized eigenvector of 
$\mUpsilon$ 
with the eigenvalue 
$\varepsilon
_{\ell}$ ($>0$).
Note that ${\cal O} =\Psi ^\dagger \Psi $ is positive semi-definite.
In other words, $ \bm{u}^\dagger_{\ell} \bm{u}_{\ell} = 1$ and  
$\mUpsilon \bm{u}_{\ell} 
= \Psi^\dagger \Psi
\bm{u}_{\ell} = \varepsilon_{\ell}\bm{u}_{\ell}$ hold.
Then, the vector $\bm{v}_{\ell} := \frac{\Psi \bm{u}_{\ell}}{\sqrt{\varepsilon_{\ell}}}$
is the normalized eigenvector of 
$H$ with the eigenvalue being $\varepsilon_{\ell}$.
One can explicitly check this as 
$\bm{v}_{\ell}^\dagger \bm{v}_{\ell}= \frac{\bm{u}^\dagger_{\ell} \Psi^\dagger\Psi \bm{u}_{\ell} }{\varepsilon_{\ell}} =\frac{\bm{u}^\dagger_{\ell} \mUpsilon \bm{u}_{\ell} }{\varepsilon_{\ell}} = 1 $
and 
$H \bm{v}_{\ell} = \frac{\Psi \Psi^\dagger  \Psi \bm{u}_{\ell}}{\sqrt{\varepsilon_\ell}} 
=  \frac{\Psi \mUpsilon \bm{u}_{\ell}}{\sqrt{\varepsilon_\ell}} 
= \sqrt{\varepsilon_\ell}  \Psi \bm{u}_{\ell}
= \varepsilon_{\ell} \bm{v}_{\ell}$. 

This relation is written in a compact form when $\det \mathcal{O} \neq 0$~\footnote{In fact, 
if the 
$\det {\cal O} =0$, 
one can also redefine $\Psi$ by reducing the number of MOs so that they are all linearly independent of each other. See Appendix~\ref{app1_1} for details.}. 
Starting from a set of $M$ orthonormalized
eigenvectors $\bm{u}_\ell $ of ${\cal O} =\Psi ^\dagger \Psi $ as
\begin{align}
  \Psi ^\dagger \Psi U =  U {\cal E},
\end{align}
where 
\begin{align}
  U = (\bm{u} _1,\cdots,\bm{u}_M)\in M(M,M),
  \end{align}
  satisfying $ U ^\dagger U =   I_M $, 
  and 
  \begin{align}
  {\cal E}   ={\rm diag}\, (\varepsilon_1,\cdots, \varepsilon _M).
\end{align}
Note that $I_M$ stands for the $M\times M$ identity matrix. 
Non-zero eigenstates $\bm{v}_\ell $, ($\ell=1,\cdots,M$) of $H=\Psi \Psi ^\dagger $ is given by
  \begin{align}
    V = (\bm{v}_1,\cdots,\bm{v}_M) 
    = \Psi U {\cal E}  ^{-1/2},\ ^\forall \varepsilon _\ell\ne 0
  \end{align}
Note that $V \in M(N,M)$, and that 
it satisfies $V^\dagger V = I_N$.
and $ \Psi \Psi^\dagger V = V {\cal E}$.
This map between the non-zero energy space is dual as
\begin{align}
  U&= \Psi^\dagger  V {\cal E}  ^{-1/2}.
\end{align}
It is further interesting to point out that this duality is associated with the square-root of the Hamiltonian~\cite{Arkinstall2017,Kremer2020,Mizoguchi2020_sq,Mizoguchi2021,Matsumoto2023,Mizoguchi2023};
we elucidate this point in Appendix~\ref{app:sqr}.

\section{Hidden Chiral symmetry \label{sec:chiral}}
We now turn to the main claim of this paper, that is, the chiral symmetry of $
{H}$.
Clearly, $
{H}$ is a positive semi-definite matrix, hence it does not preserve the chiral symmetry in a standard manner. 
However, it can preserve the hidden chiral symmetry when 
$\Psi$ satisfies certain conditions.

Suppose that the MOs are classified into two classes,
namely, we call the MO of class A (B) when 
$I =1, \cdots, M_A $ 
($I  = M_A+ 1, \cdots, M$); 
for future use, we also define 
$M_B := M - M_A$.
Then, $\Psi$ can be expressed as 
\begin{align}
  \Psi &= 
(\Psi_{\rm A}, \Psi_{\rm B} ).
  \label{eq:MO_non-overlapping}
\end{align}
Now, let us further suppose that the MOs are normalized, 
and the MOs in the same class are non-overlapping.
See also appendix \ref{app1_1}.
In other words, $\Psi_{\rm A}$
and $\Psi_{\rm B}$
satisfy
\begin{align}
\Psi^\dagger_{\rm A}\Psi_{\rm A} = \mathI_{M_A},
\hspace{1pt}
\Psi^\dagger_{\rm B}\Psi_{\rm B} = \mathI_{M_B},
\end{align}
where $\mathI_{p}$
stands for the $p \times p$ identity matrix.
Then, the overlap matrix $\mUpsilon$ becomes
\begin{align}
\mUpsilon = \begin{pmatrix}
  \Psi^\dagger_{\rm A} \\
  \Psi^\dagger_{\rm B} \\
\end{pmatrix}
(\Psi_{\rm A} , \Psi_{\rm B} ) 
= 
\begin{pmatrix}
\mathI_{M_A} & D^\dagger \\
D& \mathI_{M_B} \\
\end{pmatrix}, \label{eq:mUpsilon}
\end{align}
with $D:= \Psi^\dagger_{\rm B}\Psi_{\rm A}$.
From Eq.~(\ref{eq:mUpsilon}), we can introduce the chiral symmetry:
\begin{align}
\{ h_C,\gamma 
\} = 0,
\end{align}
with
\begin{align}
\gamma ={\rm diag}\, (I_{M_{\rm  A}}, -I_{M_{\rm  B}})=
\begin{pmatrix}
\mathI_{M_A } & {O}_{M_{\rm{A}}, M_{\rm{B}}} \\
{O}_{M_{\rm{B}}, M_{\rm{A}}} & - \mathI_{M_B },
\end{pmatrix} \label{eq:chiral_orig}
\end{align}
and
\begin{align}
h_C
= \mUpsilon-\mathI_{M}.
\end{align}
Here, ${O}_{p,q}$ 
stands for the $p \times q$ zero matrix. 
Note that $\gamma$ satisfies $\gamma^2 = \mathI_M$ and ${\rm Tr}\, \gamma =M_A-M_B$. 

Now, defining 
\begin{align}
\Gamma 
   = \Psi \gamma \Psi^\dagger,
   \label{eq:gammaprime}
  \end{align}
  and 
\begin{align}
 H_C = H-\mathI_N=\Psi \Psi ^\dagger -\mathI_N,
\end{align}
we find the following hidden chiral symmetry:
\begin{align}
\{ H_C, \Gamma\} = 0.
\label{eq:chiral_like}
\end{align}
One can explicitly check Eq.~(\ref{eq:chiral_like}) as
\begin{align}
\{ H_C, \Gamma\}
=&  (\Psi \Psi^\dagger -\mathI_N) \Psi \gamma \Psi^\dagger + \Psi \gamma \Psi^\dagger 
(\Psi \Psi^\dagger-\mathI_N) \notag \\
=&  \Psi
\left( \{   \gamma, \mUpsilon\} 
- 2 \gamma
\right) \Psi^\dagger \notag \\
=& \Psi
\{   \gamma, h_C\} 
\Psi^\dagger \notag \\
= & 0.
\end{align}
It is to be noted that 
$\Gamma$ is non-invertible if $N>M$,
i.e., $\mathrm{det}\Gamma = 0$,
because it contains $\Psi^\dagger$ 
at the rightmost [Eq.~(\ref{eq:gammaprime})].
Mathematically 
${\rm dim}\,{\rm  ker}\,  \Psi ^\dagger \ge N-M>0$, which implies $\Gamma $ has 
zero eigenvalue.
Note also that  
$(\Gamma)^2 \neq \mathI_N$.
These two features are in contrast to the conventional chiral operator.

We further describe some of the important properties of the  chiral operator $\Gamma$.
First, $\Gamma$ can be written by $\Psi_{\rm A}$ and 
$\Psi_{\rm B}$ as
\begin{align}
  \Gamma &= \Psi \gamma \Psi ^\dagger \notag 
  \\
  &= (\Psi_A,\Psi _B) \mmat{I_{M_A}}{O}{O}{-I_{M_B}}\mvec{\Psi _A ^\dagger }{\Psi _B ^\dagger }
   \notag \\
  & = \Psi _A \Psi _A ^\dagger -\Psi _B \Psi _B ^\dagger.
\end{align}
This leads to
\begin{align} 
  {\rm Tr}\, \Gamma &= {\rm Tr}\,(\Psi_A  \Psi _A ^\dagger- \Psi_B  \Psi _B ^\dagger) \notag 
  \\
  &= {\rm Tr}\,(\Psi_A ^\dagger  \Psi _A - \Psi_B ^\dagger   \Psi _B )
  \notag  \\
  &= {\rm Tr}\,(I_{M_A}-I_{M_B})
 \notag  \\
  &= M_A-M_B.
\end{align}
Second, the square of $\Gamma$ is given as
  \begin{align}
  \Gamma ^2 &= \Psi \gamma \Psi ^\dagger \Psi \gamma \Psi ^\dagger 
  \notag \\
  &= 
   \Psi \gamma (h_C+I_M)\gamma \Psi ^\dagger 
   \notag \\
    &= \Psi (-h_C+I_M)\Psi ^\dagger 
   \notag  \\
    &= \Psi ( -\Psi ^\dagger \Psi+ 2I_M)\Psi ^\dagger 
   \notag  \\
    &= -(\Psi\Psi ^\dagger )^2+2\Psi \Psi ^\dagger 
   \notag  \\
    &= -(H_C+I_N)^2 +2(H_C+I_N)
   \notag  \\
    &= I_N-H_C^2. \label{eq:sq_gamma}
  \end{align} 
This is useful for the future discussions,
  since Eq.~(\ref{eq:sq_gamma}) leads to
  the following relation, which we dub the 
  Pythagoras relation:
    \begin{align}
   \Gamma ^2+H_C^2 = I_N \label{eq:pythagoras}
\end{align}
This Pythagoras relation is consistent with the obvious commutators
due to the chiral symmetry $\acmt{\Gamma }{H_C}=0$, as 
\begin{align}
  \cmt{\Gamma}{H_C^2} &= 0,
 \end{align}
 and
 \begin{align}
  \cmt{\Gamma^2}{H_C} &= 0.
\end{align}
Also the Hermiticity of $H_C$ and $\Gamma $ implies
the positivity of their square and
\begin{align}
  -1\le \ & \xi _i \le 1,\ i=1,\cdots,N
  \\
  -1\le \ & E_i \le 1,\ i=1,\cdots,N
\end{align}
where $\xi _i$ and $E_i (=\varepsilon_i-1)$ are eigenvalues of 
$\Gamma $ 
and $H_C$, respectively. 

\section{Particle-hole symmetry and zero modes \label{sec:p-h} }
In this section, we discuss the consequence of the hidden chiral symmetry.
The spectrum of the $H_C$ is 
clear by the following observation
(let us assume  $Z\ge 0$)~\cite{Hatsugai2011}:
\begin{align}
\det\nolimits _{N}(\lambda I_N -H_C)
&= 
  \det\nolimits _{N}[(\lambda+1) I_N -\Psi \Psi ^\dagger ]
  \notag \\
  &= 
  (\lambda +1)^N\det\nolimits _{N}[ I_N -(\lambda +1)^{-1}\Psi \Psi ^\dagger ] 
  \notag \\
  &= 
  (\lambda +1)^N\det\nolimits _{M}[ I_M -(\lambda +1)^{-1}\Psi ^\dagger  \Psi  ] 
  \notag \\
  &= 
  (\lambda +1)^{N-M}\det\nolimits _{M}[ (\lambda +1)I_M-\Psi ^\dagger  \Psi  ] 
 \notag  \\
  &= 
  (\lambda +1)^{N-M}\det\nolimits _{M}[ \lambda I_M-h_C  ].
  \label{eq:det_reduce}
\end{align}
See footnote~\footnote{
Note that the general identity for any matrices 
$A_{NM}\in M(N,M), B_{MN}\in{M,N}$ \cite{Hatsugai2011}
\begin{align*}
  \det\nolimits_N (I_N+A_{NM} B_{MN}) &= \det\nolimits_M (I_M+B_{MN}A_{NM}).
\end{align*}
This obeys from a trivial one for the two square matrices $A,B\in M(N,N)$ with 
$2 N^2$ variables as their elements,    
$\det_N(I_N+AB)=\det_N A\det_N(A ^{-1} +B)=
\det_N(A ^{-1} +B)\det_N A=
\det_N(I_N+BA)$. 
Use this for $A=(A_{NM},O_{N,N-M})$ and $B=\mvec{B_{MN}}{O_{N-M,N}}$.
}.
It 
implies that the eigenvalues of $H_C$ coincide with those of $h_C$ except
$N-M$ fold degenerate (flat bands) energies at $-1$.

The chiral symmetry of $H_C$, 
$\acmt{H_C}{\Gamma }=H_C \Gamma + \Gamma H_C=0$,
implies that for any normalized eigenstate of 
$H_C$ with its energy $E$,
$  ^\forall \Phi_E$, ($\Phi_E ^\dagger  \Phi_E =1$),
$ H_C\Phi_E = \Phi_E E$, one can define
a state $\Phi^\Gamma _E$ as
\begin{align}
  \Phi^\Gamma _E &\equiv \Gamma \Phi_E=\Psi \gamma \Psi ^\dagger \Phi_E
\end{align}
It satisfies
  \begin{align}
    H_C\Phi^\Gamma _E &= H_C \Gamma \Phi _E=-\Gamma H_C \Phi_E = \Phi^\Gamma _E (-E)
\end{align}
Note that due to the Pythagoras relation, it satisfies
\begin{align}
  (\Phi^\Gamma _E) ^\dagger \Phi^\Gamma 
  _E &=\Phi_E ^\dagger \Gamma ^2 \Phi_E= 1-E^2
\end{align}

Here three cases are possible:
\begin{enumerate}
  \item[(1)] $\Phi^\Gamma _E =0$ (its norm is zero). $E=\pm 1$.
\begin{enumerate}
  \item [(1-1)]$E=+1$, $\Gamma \Phi_{E=+1}\ne 0$.
  \item [(1-2)]$E=-1$, $\Gamma \Phi_{E=-1}= 0$. 

    $\Phi_{E=-1}$ belongs to the "flat band".
\end{enumerate}
  \item[(2)] $E\ne\pm 1,0$, $\Phi^\Gamma _E $ is a non-vanishing eigenstate of  the energy $-E$. The states $\Phi_{E}$ and $\Phi_{-E}$ are paired.

  \item[(3)] $E=0$, $H_C\Phi_{E=0}=0$. Due to the Pythagoras relation, 
    $\Gamma^2 \Phi_{E=0}=\Phi_{E=0}$.
\end{enumerate}
For the case (1), due to the Pythagoras relation, $E^2=1$.
If $E=-1$, $\Phi_{E=-1}$ belongs to the "flat band".
The other case is also possible, $\Phi_{E=1}$, which 
induces $\Phi^\Gamma _{E=1}=0$.

For the case (2), two states $\Phi_E$ and $\Phi^\Gamma _E$ are 
paired with the pair energies $E$ and $-E$.
Note that $(\Phi^\Gamma _E)^\Gamma =(1-E^2)\Phi_E$ due to 
the Pythagoras relation. It guarantees that both of the pair are 
of finite norm.
Also if two degenerate states $\Phi_{E}^1$ and $\Phi_{E}^2$ are
orthogonal, $(\Phi_{E}^1) ^\dagger \Phi_{E}^2 =0$, 
the partners of the pair
satisfy 
 $({\Phi^\Gamma _{E}}^1) ^\dagger {\Phi^\Gamma _{E}}^2 =
 (1-E^2)
(\Phi_{E}^1) ^\dagger \Phi_{E}^2 =0$.
Then the degeneracy of the states at 
$\pm E$ are the same unless $E=\pm 1$. 

For the  case (3), the zero mode, $\psi_{E=0}$ is special.
Since $\Gamma \Phi_{E=0}$ also belongs to the zero mode,
it is expanded by the eigenstates of the zero modes, 
which are 
complete. It implies that one can 
diagonalize $\Gamma$ within 
the subspace of the zero modes, that is, 
we may assume the zero modes
are eigenstates of the chiral operator $\Gamma $
with its eigenvalue (chirality) $\pm 1$ due to the Pythagoras relation as
\begin{align}
  \Gamma\Phi^+_{i_+,E=0} &= (+1)\Phi^+_{i_+,E=0},\ i_+=1,\cdots,N_+,
  \\
  \Gamma\Phi^-_{i_-,E=0} &= (-1)\Phi^-_{i_-,E=0},\ i_-=1,\cdots,N_-,
\end{align}
where $N_\pm$ are the number of the zero modes with chiralities $\pm 1$.
Here let us write an eigenstate of $\Gamma $ with its eigenvalue $\xi$ 
as $\bm{\varphi}_{\xi}$.
Then due to the chiral symmetry $\acmt{H_C}{\Gamma} =0$, 
similar to the above discussion, 
eigenstates of $\Gamma $ are paired.
Namely, for 
$\varphi_{\xi}$ satisfying 
$\Gamma \bm{\varphi}_{\xi} = \xi \bm{\varphi}_{\xi}$ with $\xi \neq 1$,
we find
\begin{align}
  \Gamma \bm{\varphi}^{H_C}_{\xi } = -\xi \bm{\varphi}^{H_C}_{\xi},
\end{align} 
where $\bm{\varphi}^{H_C}_\xi =H_C \bm{\varphi}_\xi$.
This pairing of the chiral operator (chiral pair)
is up to the degeneracy unless $\xi\ne \pm 1$, which 
implies the cancelation of the chiralities (for ${\rm Tr}\, \Gamma $) 
of the chiral operator except $\xi =\pm 1$.
It results in an important constraint between the number of the
zero modes of the $H_C$ and the chiral operator~\cite{Sutherland1986} 
\begin{align}
  N_+-N_- &=
  {\rm Tr}\,\Gamma =
 M_A-M_B.
  \label{eq:index}
\end{align}
\begin{figure}[!tb]
\begin{center}
\includegraphics[clip,width = \linewidth]{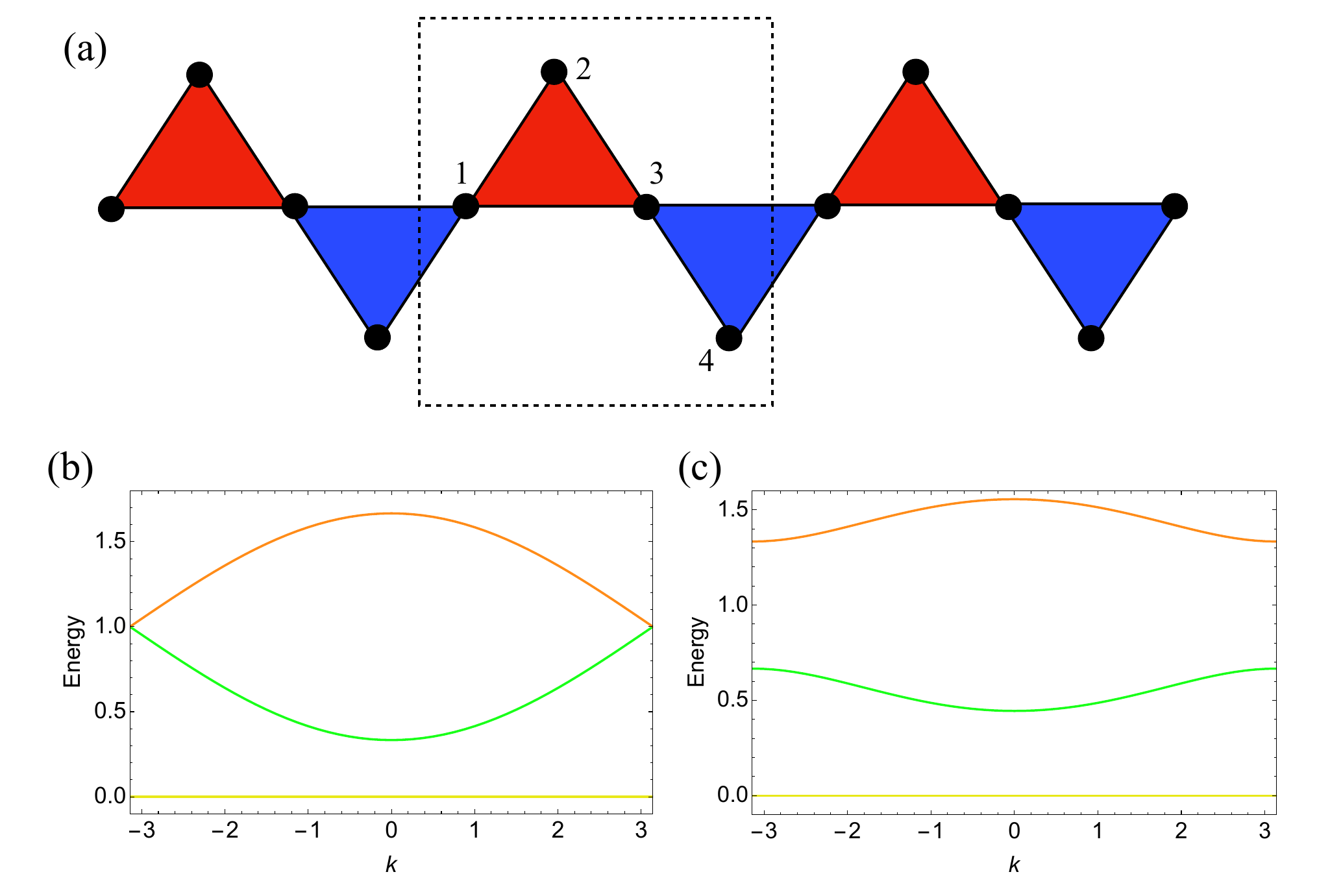}
\vspace{-10pt}
\caption{
(a) Modified saw-tooth lattice.
The panels (b) and (c) are the band structures for $(\alpha_1, \alpha_2, \alpha_3,\beta_1, \beta_2, \beta_3) = $
(b) $\left(\frac{1}{\sqrt{3}},\frac{1}{\sqrt{3}},\frac{1}{\sqrt{3}},\frac{1}{\sqrt{3}},\frac{1}{\sqrt{3}},\frac{1}{\sqrt{3}}  \right)$
and 
(c) $\left(\frac{2}{3},\frac{2}{3},\frac{1}{3},\frac{1}{3} ,\frac{2}{3},\frac{2}{3}  \right)$ .
}
\label{fig:ST}
\end{center}
\vspace{-10pt}
\end{figure}
Since the zero modes of $H_C$ and $h_C$ 
correspond to the energy $\varepsilon =+1$ in the
section \ref{sec:mo}, the duality is safely applied as
\begin{align}
  \Phi_{i\pm,E=0}^\pm &= \Psi \bm{\phi}_{i\pm,E=0},
  \end{align}
  and 
  \begin{align}
  \bm{\phi}_{i\pm,E=0}^\pm &= \Psi ^\dagger \Phi_{i\pm,E=0},
\end{align}
where
  $\phi_{i\pm,E=0} $ is a zero mode of $h_C$ with chirality $\pm 1$ of
  $\gamma $ as
\begin{align}
  h_C  \bm{\phi}_{i\pm,E=0} = 0,
  \hspace{1pt}
  \gamma  \bm{\phi}_{i\pm,E=0} = \pm  \bm{\phi}_{i\pm,E=0}.
\end{align}
This map implies the direct proof of 
the Eq.~(\ref{eq:index}); 
see Appendix~\ref{app1_2}.

\begin{figure}[!tb]
\begin{center}
\includegraphics[clip,width = \linewidth]{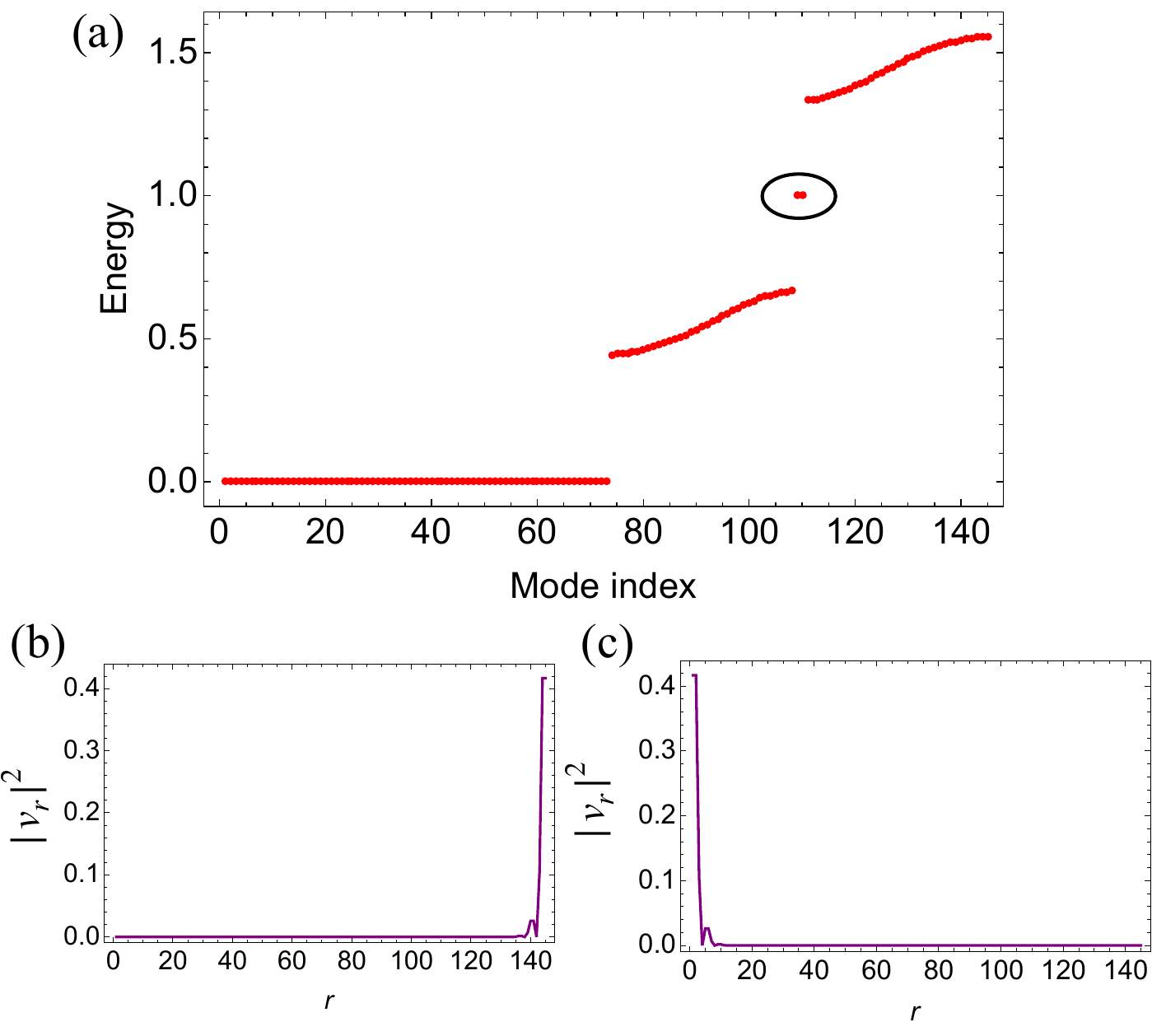}
\vspace{-10pt}
\caption{
(a) Energy spectrum for the modified saw-tooth lattice with $(\alpha_1, \alpha_2, \alpha_3,\beta_1, \beta_2, \beta_3) =\left(\frac{2}{3},\frac{2}{3},\frac{1}{3},\frac{1}{3},\frac{2}{3},\frac{2}{3}  \right)$.
The in-gap states are encircled by a black ellipse. 
The panels (b) and (c) shows the particle density 
for the two in-gap states.
Note that the horizontal axis is $r = 4R +a$.
}
\label{fig:ST_open}
\end{center}
\vspace{-10pt}
\end{figure}

In the following, 
we show two examples of the MO models preserving 
this chiral symmetry,
and discuss their characteristic features arising from this symmetry.

\begin{figure}[!tb]
\begin{center}
\includegraphics[clip,width = \linewidth]{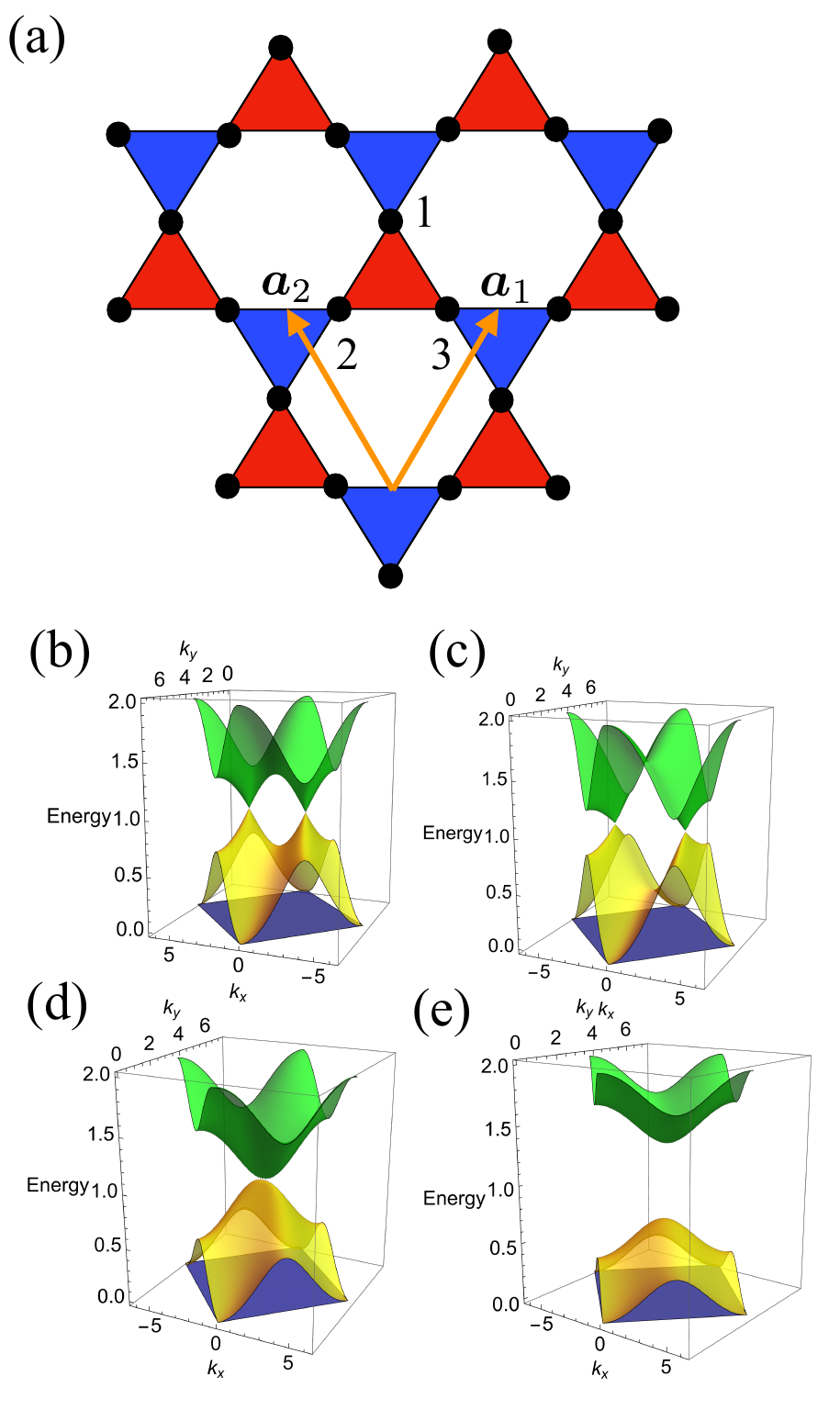}
\vspace{-10pt}
\caption{
(a) Kagome lattice. 
The panels (b)-(e) are the band structures for 
$(\alpha_1,\alpha_2,\alpha_3,\beta_1,\beta_2, \beta_3)=$
(b) $\left(\frac{1}{\sqrt{3}},\frac{1}{\sqrt{3}},\frac{1}{\sqrt{3}},\frac{1}{\sqrt{3}},\frac{1}{\sqrt{3}},\frac{1}{\sqrt{3}}  \right)$,
(c) $\left(\frac{1}{\sqrt{5}},\frac{\sqrt{2}}{\sqrt{5}},\frac{\sqrt{2}}{\sqrt{5}},\frac{1}{\sqrt{5}},\frac{\sqrt{2}}{\sqrt{5}},\frac{\sqrt{2}}{\sqrt{5}}  \right)$,
(d) $\left(\frac{1}{\sqrt{2}},\frac{1}{2},\frac{1}{2},\frac{1}{\sqrt{2}},\frac{1}{2},\frac{1}{2}  \right)$,
and (e)
$\left(\frac{\sqrt{2}}{\sqrt{3}},\frac{1}{\sqrt{6}},\frac{1}{\sqrt{6}},\frac{\sqrt{2}}{\sqrt{3}},\frac{1}{\sqrt{6}},\frac{1}{\sqrt{6}}  \right)$.
}
\label{fig:kagome}
\end{center}
\vspace{-10pt}
\end{figure}

\section{Example 1: Modified saw-tooth lattice \label{sec:sawtooth}}
We first consider the modified saw-tooth lattice [Fig.~\ref{fig:ST}(a)]
where the upward triangles and the downward triangles are connected in 
an alternating manner. 
We label the unit cell 
[denoted by a broken rectangle in Fig.~\ref{fig:ST}(a)] 
by $R = 0, \cdots, L-1$ 
and the sublattices by $a = 1,2,3,4$.
Clearly, the lattice itself is non-bipartite and the chiral symmetry is not preserved in the ordinal sense. 

On this lattice, 
we define two 
species of MOs:
\begin{align}
C^\dagger_{R,\bigtriangleup} 
=\alpha_1 c^\dagger_{R,1} +  \alpha_2 c^\dagger_{R,2}  + \alpha_3 c^\dagger_{R,3},
\end{align}
and
\begin{align}
C^\dagger_{R,\bigtriangledown} 
=\beta_1 c^\dagger_{R,3}  + \beta_2 c^\dagger_{R,4} +\beta_3 c^\dagger_{R+1,1}.
\end{align}
In the terminology of Sec.~\ref{sec:chiral},
$\bigtriangleup$ corresponds to A, and 
$\bigtriangledown$ to B.
The normalization condition for 
$C^\dagger_{R,\bigtriangleup}$
and $C^\dagger_{R,\bigtriangledown}$
reads,
respectively, $\sum_{j=1}^{3} |\alpha_j|^2 = 1$
and $\sum_{j=1}^{3} |\beta_j|^2 = 1$.
In that case, 
the condition of 
Eq.~(\ref{eq:MO_non-overlapping})
is satisfied. 
Note that $N= 4L$ and $M=2L$. 

As the model is translationally invariant, 
we can adopt 
the band representation when imposing the periodic boundary condition. 
By performing the Fourier transform, one obtains
the momentum-space representation of $\Psi$, which we 
write $\Psi_k$, as
\begin{align}
\Psi_{k} = 
\begin{pmatrix}
\bm{\psi}_{k, \bigtriangleup} & 
\bm{\psi}_{k, \bigtriangledown} \\
\end{pmatrix},
\end{align}
with 
\begin{align}
\psi_{k, \bigtriangleup} = (\alpha_1,\alpha_2,\alpha_3,0)^{\rm T},
\end{align}
and 
\begin{align}
\psi_{k, \bigtriangledown}
= (\beta_3 e^{-ik}, 0, \beta_1, \beta_2)^{\rm T}.
\end{align}
Using these, we define the momentum-space 
representation of $H$, 
which we write $H_k$,
and the band structure corresponds to the eigenvalues of $H_k$.
Figures~\ref{fig:ST}(b) and \ref{fig:ST}(c)
are the band structures 
for two representative parameters.
We indeed see zero-energy flat band, 
which is doubly degenerate. 
This coincides with the aforementioned counting of the zero modes.
We also see that the finite-energy modes 
are symmetric with respect to $\varepsilon = 1$.
In fact, 
$h_C$ for this model corresponds to 
the tight-binding model on a chain with alternating nearest-neighbor hoppings [so-called the Su-Schriffer-Heeger (SSH) model~\cite{Su1979}].

We next show the emergence 
of the topological edge mode, 
which is a consequence of the chiral symmetry. 
As is well known, the SSH model is one of the representative examples of the topological insulator of the chiral-symmetric class~\cite{Schnyder2008,Kitaev2009,Ryu2010,Topo_book}, so we expect that the topological phase is inherited to the modified saw-tooth lattice model due to the chiral symmetry.
We consider the system under the open boundary condition, hosting $L$ upward triangles and $L$ downward triangles: 
the number of sites is $N= 4L+1$ and the total number of MOs is $M= 2L$. 
In Fig.~\ref{fig:ST_open}(a),
we plot the energy spectrum for the system with $L= 36$.
We see the degenerate zero modes, whose degeneracy is 
$2L+1$. 
We also see that there exist two in-gap states at $\varepsilon \sim 1$~\footnote{Note that the energy is not exactly equal to 1 due to the finite-size effect.}.
The particle density distributions for these two modes are shown in Figs.~\ref{fig:ST_open}(b) and \ref{fig:ST_open}(c).
We see that one is localized on the right edge and the other is localized on the left edge.
These finite-energy edge modes 
can be viewed as a topological edge modes protected 
by the hidden chiral symmetry $\Gamma$.

\begin{figure}[!tb]
\begin{center}
\includegraphics[clip,width = \linewidth]{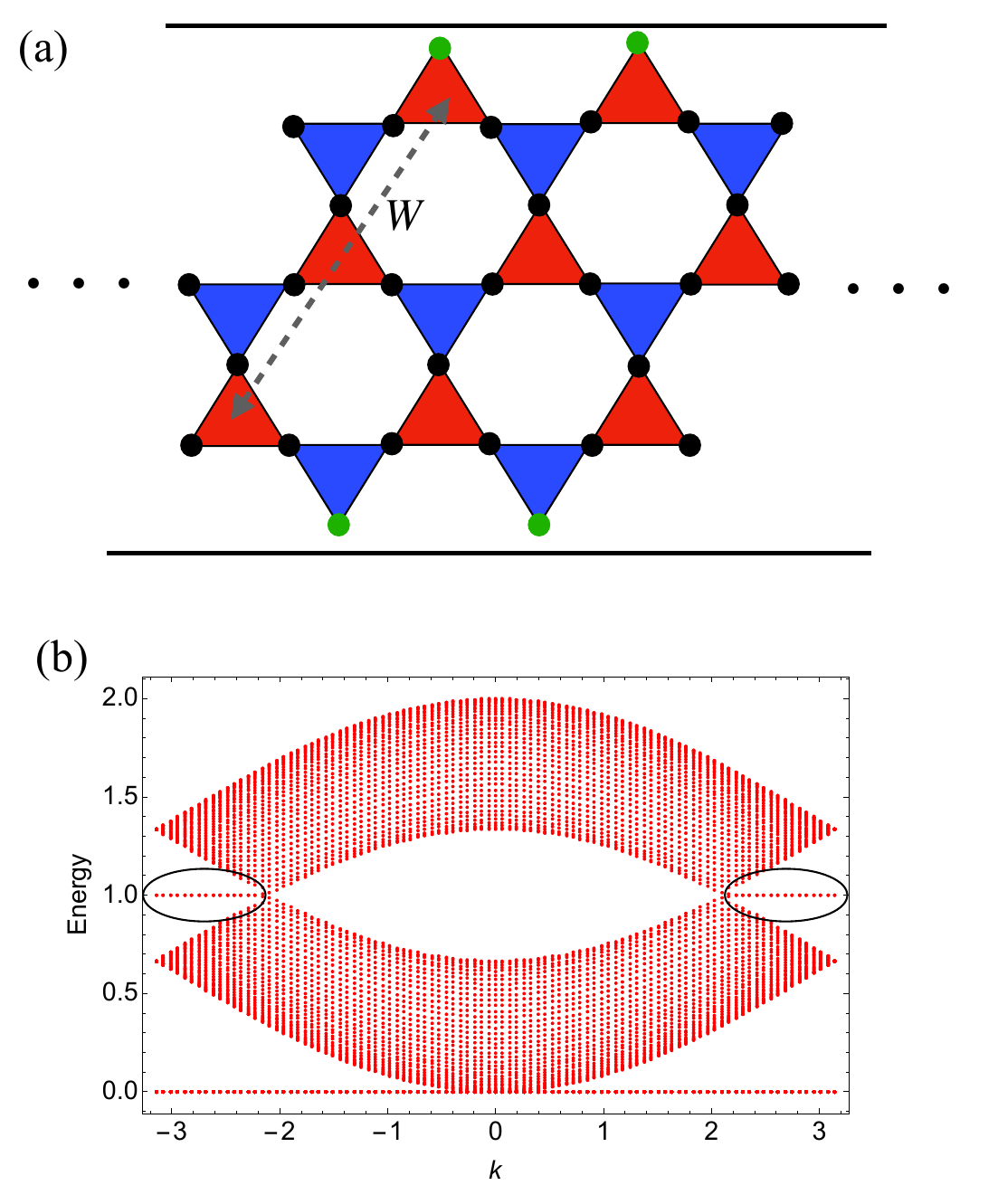}
\vspace{-10pt}
\caption{
(a) Finite strip of the kagome lattice. 
(b) The energy spectrum 
for $(\alpha_1,\alpha_2,\alpha_3,\beta_1,\beta_2, \beta_3)=\left(\frac{1}{\sqrt{3}},\frac{1}{\sqrt{3}},\frac{1}{\sqrt{3}},\frac{1}{\sqrt{3}},\frac{1}{\sqrt{3}},\frac{1}{\sqrt{3}}  \right)$ and $W = 36$.
The flat edge modes are encircled by the black ellipses. 
}
\label{fig:kagome_open}
\end{center}
\vspace{-10pt}
\end{figure}

\section{Example 2: Kagome lattice \label{sec:kagome}}
The second example is the kagome lattice, which is the corner-sharing network of the triangles
[Fig.~\ref{fig:kagome}(a)]. 
We label the unit cell 
by $\bm{R} = n_1 \bm{a}_1 + n_2 \bm{a}_2$
($n_{1/2} = 0, \cdots, L-1$)
and the sublattices by $a = 1,2,3$. 
For this model, we define two species of 
MOs:
\begin{align}
C^\dagger_{\bm{R},\bigtriangleup}=
\alpha_1 c^\dagger_{\bm{R},1} +
\alpha_2 c^\dagger_{\bm{R},2}+ 
\alpha_3 c^\dagger_{\bm{R},3},
\end{align}
and 
\begin{align}
C^\dagger_{\bm{R},\bigtriangledown}=
\beta_1 c^\dagger_{\bm{R},1} +
\beta_2 c^\dagger_{\bm{R}+\bm{a}_1,2}+ 
\beta_3 c^\dagger_{\bm{R}+\bm{a}_2,3},
\end{align}
which again corresponds to the MOs 
for the upward and downward triangles, respectively.
The conditions for normalization to be assigned for $\alpha$s' and $\beta$s' 
are the same as that for the modified saw-tooth lattice.
Note that, in this model, 
$N= 3L^2$ and $M= 2L^2$.

We first argue the band structure under the periodic boundary condition.
We note that the momentum-space representation of $\Psi_{\bm{k}}$ in this case is
\begin{align}
\Psi_{\bm{k}} = 
\begin{pmatrix}
\bm{\psi}_{\bm{k}, \bigtriangleup} & 
\bm{\psi}_{\bm{k}, \bigtriangledown} \\
\end{pmatrix}, \label{eq:mo_kagome_k}
\end{align}
with 
\begin{align}
\psi_{\bm{k}, \bigtriangleup} = (\alpha_1,\alpha_2,\alpha_3)^{\rm T},
\end{align}
and 
\begin{align}
\psi_{\bm{k}, \bigtriangledown}
= (\beta_1, \beta_2e^{-i\bm{k}\cdot \bm{a}_1},\beta_3 e^{-i\bm{k}\cdot \bm{a}_2})^{\rm T}.
\end{align}

Figures~\ref{fig:kagome}(b)-\ref{fig:kagome}(e) are the band structures for some sets of parameters.
We see the zero-energy flat band and the symmetric finite-energy bands around $\varepsilon = 1$.
In Fig.~\ref{fig:kagome}(b), where $\alpha$s' and $\beta$s' are all equal to $1/\sqrt{3}$, the dispersion of the finite-energy bands 
is, as is well-known~\cite{Hatsugai2011,Essafi2017,MizoguchiUdagawa2019}, 
the same as that of the honeycomb-lattice tight-binding model (or graphene) which hosts the Dirac cones at $K$ and $K^\prime$ points in the Brillouin zone.
The pair of Dirac cones implies the Fermion doubling~\cite{Nielsen1981,Nielsen1981_2}.  
The zero modes at these points can be obtained explicitly. To be concrete, 
at the K and K$^\prime$ points,
we have
\begin{align} 
      \psi_{K, \bigtriangleup} =\psi_{K^\prime, \bigtriangleup} 
      = \frac{1}{\sqrt{3}}\mvecthree{1}{1}{1},
      \end{align}
      and
      \begin{align}
      \psi_{K, \bigtriangledown} 
      =  \psi^\ast_{K^\prime, \bigtriangledown} 
      =\frac{1}{\sqrt{3}}\mvecthree{1}{\omega^2 }{\omega},
      \end{align}
      with $\omega := e^{i \frac{2\pi}{3}}$.
    As $h_{K/K ^\prime, C}=0$, the zero modes of $h_{K/K ^\prime, C}$ with fixed chiralities,
    $h_{\rm K, K ^\prime, C} \bm{\phi}_\pm=0$ and      
    $\gamma \bm{\phi}^\pm=\pm \bm{\phi}^\pm$,
    are given by
    \begin{align}
      \bm{\phi}^+ = \mvec{1}{0}, \hspace{1pt}
      \bm{\phi}^- = \mvec{0}{1}.
    \end{align}
    The corresponding zero modes of 
    $H_{K/K^\prime,C}$ with fixed chiralities
    are
    \begin{align} 
      \Phi^+_{K/K^\prime} = \Psi_{K/K^\prime} \bm{\phi}^+=\psi_{K/K^\prime,\bigtriangleup},
    \end{align}
    and 
    \begin{align}
      \Phi^-_{K/K^\prime} = \Psi_{K/K^\prime} \bm{\phi}^-=\psi_{K/K^\prime,\bigtriangledown},
    \end{align} 
    where we can directly check 
    \begin{align}
      H_{K/K^\prime,C} \Phi^\pm_{K/K^\prime} = 0,
      \end{align}
      and \begin{align}
      \Gamma_{K/K^\prime} \Phi^\pm_{K/K^\prime} = \pm \Phi^\pm_{K/K^\prime}.
    \end{align}

The pair of Dirac cones is robust against 
changing $\alpha$s'
and $\beta$s' from $1/\sqrt{3}$ 
(and can be complex numbers as well)
unless the deviation is too large. 
This topological stability arises from the fact 
that the Dirac cones correspond to 
those of $h_C$, as we have mentioned in Sec.~\ref{sec:p-h} and Sec.~\ref{app1_2}.
In this sense, 
it is protected by the hidden chiral symmetry.
Here, we demonstrate it. In Figs.~\ref{fig:kagome}(c)-\ref{fig:kagome}(e),
we choose the parameters such that the finite-energy bands become the same as those of the anisotropic honeycomb model~\cite{Hasegawa2006,Montambaux2009,Montambaux2009_2,Hasegawa2012}, where one of the three directions of the bonds have the different hopping parameter from the other two (also known as the effective model of the phosphorene~\cite{Ezawa2014,Ezawa2018}).
This modulation does not break the chiral symmetry, hence the Dirac cones 
are robust against the change of the parameters [Fig.~\ref{fig:kagome}(c)].
By further changing the parameters, they merge at the same momentum and form the semi-Dirac dispersion [Fig~\ref{fig:kagome}(d)], and gapped out in a pairwise manner [Fig.~\ref{fig:kagome}(e)].
This evolution of the Dirac cones is exactly the same as that for the  anisotropic honeycomb model~\cite{Hasegawa2006,Montambaux2009,Montambaux2009_2,Hasegawa2012}.

We note that, for Figs.~\ref{fig:kagome}(b)-\ref{fig:kagome}(e),
the lower dispersive band touches the flat band at $\bm{k}=\bm{0}$.
This means that the degeneracy of the zero-energy modes 
is $L^2 + 1$.
We also see that, at the same momentum, 
the eigenenergy is $\varepsilon = 2$.
This coincides with the argument presented in Sec.~\ref{sec:p-h}.

Next, we address the topological edge modes resulting from the chiral symmetry.
To this end, we consider finite-width strip of the kagome lattice, shown in Fig.~\ref{fig:kagome_open}(a).
We denote the width of the strip by $W$, which is equal to the number of upward triangles along the gray dashed arrow in Fig.~\ref{fig:kagome_open}(a).
We assign the periodic boundary condition in the horizontal direction. 
In Fig.~\ref{fig:kagome_open}(b), we plot the energy spectrum 
of the kagome-lattice strip with $W= 36$ with $\alpha_j = \beta_j = \frac{1}{\sqrt{3}}$ for $j=1,2,3$. 
We see that the flat edge modes exist
for $\frac{2\pi}{3} \leq |k| \leq \pi$.
In fact, $h_C$ 
under this boundary condition corresponds to the graphene under the zigzag edge, 
which is the origin of the flat edge modes.
As the flat edge modes of graphene are topologically protected due to the chiral symmetry~\cite{Ryu2002},
the flat edge modes of the 
kagome-lattice strip can 
also be understood as 
a consequence of the hidden chiral symmetry.
Note that, in the original basis 
of the kagome sites, 
the Hamiltonian contains 
the nearest-neighbor hopping with the amplitude 
$\frac{1}{3}$ and the on-site potential with the amplitude $\frac{2}{3}$, except for the sites along the top and bottom boundaries [the sites colored in green in Fig.~\ref{fig:kagome_open}(a)]; 
for these sites, 
the on-site potential is $\frac{1}{3}$.
This indicates that the tuning of the on-site potential on the edge enables us to realize the graphene-like flat-edge modes even for the kagome lattice.
\begin{figure}[b]
\begin{center}
\includegraphics[clip,width = \linewidth]{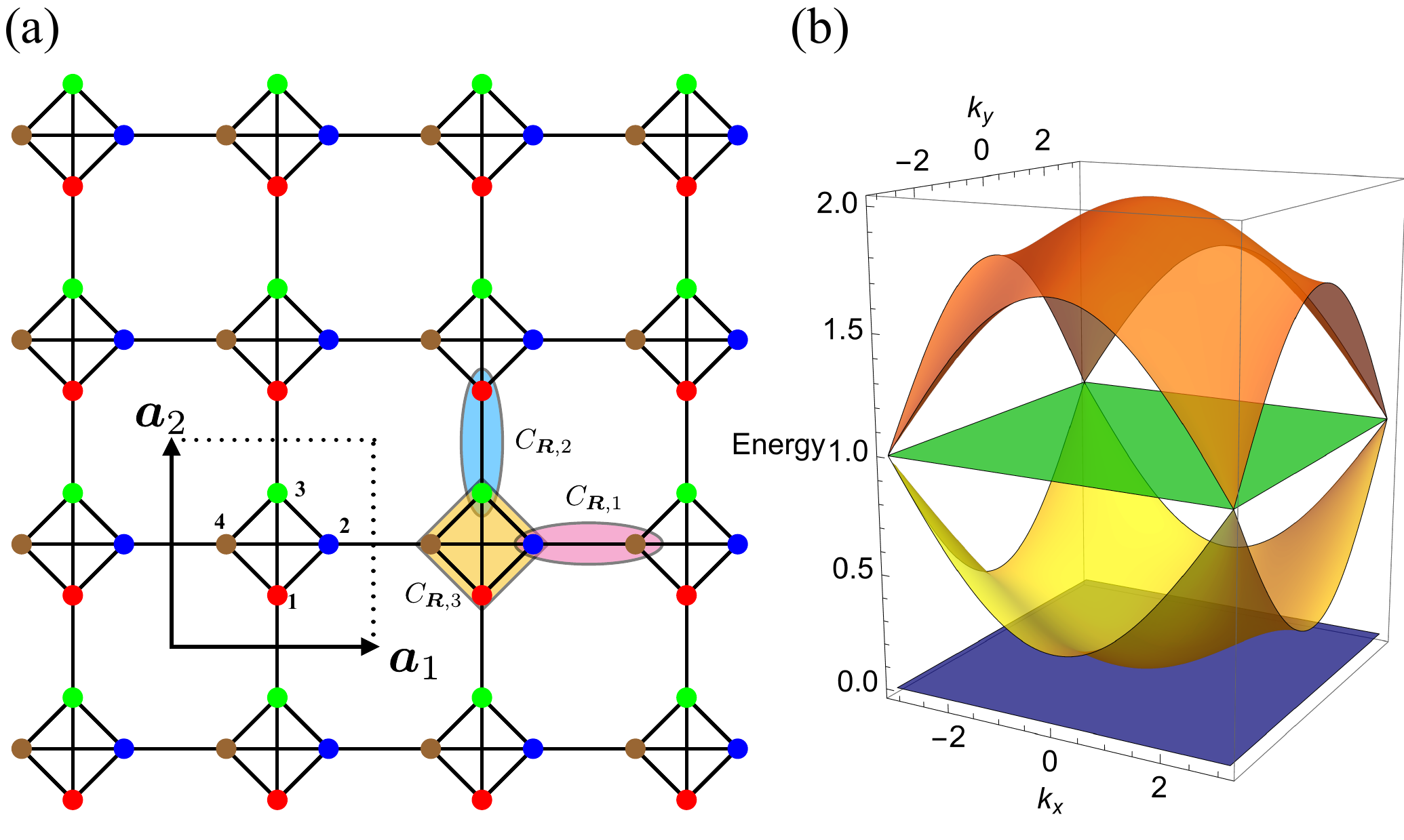}
\vspace{-10pt}
\caption{
(a) Square-octagon lattice. 
(b) Band structures for 
$(\alpha_1,\alpha_2,\alpha_3,\alpha_4,\beta_1,\beta_2, \beta_3,\beta_4)=
\left(\frac{1}{\sqrt{2}},\frac{1}{\sqrt{2}},\frac{1}{\sqrt{2}},\frac{1}{\sqrt{2}},
\frac{1}{2},\frac{1}{2},\frac{1}{2},\frac{1}{2}
\right)$.
}
\label{fig:sqo}
\end{center}
\vspace{-10pt}
\end{figure}

We close this section by remarking that, 
for the kagome lattice model, 
the eigenvalues of the chiral operator $\Gamma$
has characteristic features that originate from the 
properties we discussed in Secs.~\ref{sec:chiral}
and \ref{sec:p-h}.
We address this point in Appendix~\ref{app:remark}.

\section{Example 3: Square-octagaon lattice \label{sec:sqo}}
As the third example, we consider the square-octagon lattice model [Fig.~\ref{fig:sqo}(a)].
The characteristics of this model 
is that the number of MOs belonging to the class A is not equal to that to the class B.
We label the unit cell 
by $\bm{R} = n_1 \bm{a}_1 + n_2 \bm{a}_2$
($n_{1/2} = 0, \cdots, L-1$)
and the sublattices by $a = 1,2,3,4$. 
For this model, 
we define three species of 
MOs:
\begin{align}
C^\dagger_{\bm{R},1}=
\alpha_1 c^\dagger_{\bm{R},2} +
\alpha_2 c^\dagger_{\bm{R}+\bm{a}_1,4},
\end{align}
\begin{align}
C^\dagger_{\bm{R},2}=
\alpha_1 c^\dagger_{\bm{R},3} +
\alpha_2 c^\dagger_{\bm{R}+\bm{a}_2,1}
\end{align}
and 
\begin{align}
C^\dagger_{\bm{R},3}=
\beta_1 c^\dagger_{\bm{R},1} +
\beta_2 c^\dagger_{\bm{R},2}+ 
\beta_3 c^\dagger_{\bm{R},3}+
\beta_3 c^\dagger_{\bm{R},4}.
\end{align}
See Fig.~\ref{fig:sqo}(a) for the schematic figure of the MOs. 
The normalization condition is $|\alpha_{1}|^2 + |\alpha_{2}|^2 = |\alpha_{3}|^2 + |\alpha_{4}|^2
= \sum_{a=1}^{4} |\beta_{a}|^2 = 1$.
We regard $C^\dagger_{\bm{R},1}$ and $C^\dagger_{\bm{R},2}$ are in the class A
and $C^\dagger_{\bm{R},3}$ is in the class B.
Then we have $M_A= 2L^2$ and $M_B = L^2$,
so this model is an example of 
$M_A \neq M_B$.
Note that $h_C$ for this model corresponds to the tight-binding model on a Lieb lattice~\cite{Lieb1989}. 

Figure~\ref{fig:sqo}(b) shows the band structure for the model (the parameters are shown in the caption). 
Remarkably, there exist two flat bands;
one is at $\varepsilon = 0$ and the other is at $\varepsilon = 1$.  
The former is obtained by the generic argument of the MO representation.
Meanwhile, 
the latter corresponds to the zero-modes for $H_C$.
In fact, it
originates from 
the fact that $M_A \neq M_B$,
namely, Eq.~(\ref{eq:index}) 
guarantees the existence of the zero modes for $H_C$ when $M_A \neq M_B$.

\section{Summary and discussions \label{sec:summary}}
We have presented the method of constructing the flat-band models that preserve the hidden chiral symmetry of the finite-energy modes. 
Our construction relies on the MO representation, by which we can guarantee the existence of the zero-energy flat bands. 
The main claim of this work is that 
the MO models can preserve the chiral symmetry, 
despite being defined on the non-bipartite lattice, 
when the MOs satisfy the ``separatable" and ``non-overlapping" conditions. 
Importantly, the chiral operator $\Gamma$ 
for the present model has the distinctive features from the conventional ones, as the chiral operator itself is constructed by the same MOs as the Hamiltonian.
One of its manifestations is 
the Pythagoras relation of Eq.~(\ref{eq:pythagoras}),
which ties the eigenvalues of $H_C$ with those of the $\Gamma$.
As a consequence of the chiral symmetry, 
the band structures of the finite-energy bands 
are particle-hole symmetric around $\varepsilon = 1$. 
We have presented two concrete examples, namely, 
the modified saw-tooth lattice model and the kagome lattice model,
and have discussed their characteristic band structures 
as well as the topological boundary modes. 
We additionally show the example of the square-octagon lattice as an example of $M_A \neq M_B$.
The characteristics of this type of model is that there exist two flat bands with different energies,
one of which originates from the hidden chiral symmetry under the condisition of $M_A \neq M_B$. 

Our construction applies to higher-dimensional systems, such as the pyrochlore lattice. 
It also applies 
to the random systems, too. 
Actually, the random U(1) MO model which we have introduced in the previous work~\cite{Mizoguchi2023} is one of the examples of this class of model. 
Further studies on the spatially non-uniform models,
such as revealing the characteristics of the localization phenomena, will be worth considering
in the future work.   

As for the experimental realization, 
the three examples we present in this paper consist of 
finite-ranged hopping only, 
but the hopping amplitude 
and the on-site potentials are not independent.
Considering the fact that 
the fine-tuning is needed, 
we think that the electric circuit~\cite{Albert2015,Lee2018} will be a good platform of realizing the topological boundary modes shown in this paper.

\acknowledgments
We thank H. Tsunetsugu and A. Furusaki for useful comments on our previous work (Ref.~\onlinecite{Mizoguchi2023}).
This work is supported by 
JSPS KAKENHI, Grant No.~JP23K03243 (T.M.) and 
No.~23K25788 (Y.H.),
and by JST-CREST Grant No.~JPMJCR19T1.
T.M. and Y.H. contributed equally to this work.

\appendix
\section{Molecular orbitals for one-particle states \label{app1}}
  \subsection{Generic framework \label{app1_1}}
  The meaning of the molecular orbital $\Psi$ is clear by restricting 
  the discussion to
  the one-particle states~\cite{Hatsugai2021}. 
  Here we re-formulate the MO representation 
  by using the language of the single-particle states. 
  Let us start with a standard AO
  $|i \rangle $ which is localized at the atomic site $i$
  \begin{align}
    | i \rangle &= c_i ^\dagger |0 \rangle ,\ i=1,\cdots,N, 
  \end{align} 
  where
  $c_i$ is a annihilation operator
  of a canonical fermion labeled at the atomic site $i$,
  $\acmt{c_i}{c_j ^\dagger } = \delta _{ij}$ and
  $|0 \rangle $ is their vacuum $^\forall c_i | 0 \rangle =0$.
  These AOs are orthonrmalized 
  $ \langle i|j \rangle = \delta _{ij}$,
  \begin{align}
    \langle {\cal A}| {\cal A} \rangle &= I_N,
    \notag \\
    |{\cal A} \rangle  &= (|1 \rangle ,\cdots,|N \rangle )= \bm{c}^\dagger |0 \rangle,  
    \notag \\
    \bm{c}^\dagger &= (c_1 ^\dagger ,\cdots,c_N ^\dagger ).
  \end{align} 
  The molecular orbital $|I )$, ($I=1,\cdots,M$)
  is spanned by the atomic orbitals as
  \begin{align}
    | I ) &=  \sum_{i=1}^N| i \rangle \psi  _{i,I}=|{\cal A} \rangle \psi_I, 
  \end{align}
  where
  $\psi_I$ is an $N$-dimensional column vector of the $I$-th molecular orbital.
  and $| {\cal A} \rangle $ is a set of
  orthonormalized atomic basis,
  The fermion operator of the molecular orbital $C_I$ is defined by
  \begin{align} 
    | I ) &= C_I ^\dagger |0 \rangle ,\ I=1,\cdots,M,
    \notag \\
    C_I ^\dagger  &= \bm{c} ^\dagger \psi_I.
  \end{align} 
  Then the set of the MOs is written as
  \begin{align}
    |{\cal M} ) &= (|1),\cdots,|M))= |{\cal A}  \rangle \Psi,
   \notag \\
    \Psi &= (\psi_1,\cdots, \psi_M)\in M(N,M),
   \notag \\
    \bm{C}  ^\dagger &= \bm{c}  ^\dagger \Psi.
  \end{align}
  Here let us consider a one-particle
  Hamiltonian 
  written by the $M$ molecular orbitals (MOs) as~\cite{Hatsugai2011,Hatsugai2015,Mizoguchi2019,Mizoguchi2020,Hatsugai2021,Mizoguchi2021_skagome,Mizoguchi2022}
  \begin{align}
    {\cal H}_1     &= \sum_{I,J}^M| I )h_{I,J}(J|=|{\cal M} ) {h} ({\cal M} |
    =|{\cal A} \rangle {H} \langle{\cal A} |,
   \end{align}
   with
   \begin{align}
    H    &= \Psi h \Psi ^\dagger,
  \end{align}
  where $H\in M(N,N)$, $ {h} \in M(M,M)$ 
  and $ h_{i,j}=\{ h \}_{i,j}$, 
  is a hopping between the molecular orbitals $|i)$ and $|j)$.

  The overlap matrix of the MOs is defined as
  \begin{align}
    {\cal O} &= ({\cal M} |{\cal M} )=\Psi ^\dagger \Psi \in M(M,M).
  \end{align} 
  This overlap matrix
  ${\cal O} $ is semi-positive definite, that is, its 
  eigenvalue $o$ of ${\cal O} $ is positive or zero.
  It is clear by  
  $o = \bm{u} ^\dagger {\cal O} \bm{u}= \| \Psi \bm{u}\|^2$ 
  where $\bm{u}$ is the normalized eigenvector of ${\cal O} $.

  Although the number of the MOs
  can be $M\geq N$ or $M<N$, 
  the number of the independent 
  MOs $\bar M $ 
  does not exceed that 
  of the total atomic orbitals. 
  Technically it is written as 
  $\bar M={\rm rank}\, \Psi \le N$ since $\Psi \in M(N,M)$. 
  Here
  $\text{rank}\, \Psi$ is the largest dimension of the nonzero minor determinant of
  the order $k$, $\det \Delta_{i_1,\cdots,i_k;j_1,\cdots,j_k}$,
  $(\Delta_{i_1,\cdots,i_k;j_1,\cdots,j_k})_{a,b} =\Psi_{i_a,j_b}$,
  ($1\le a,b\le k$).

  Let us assume some of the eigenstates, $o=0$ and  $\Psi \bm{v} =0$.
  Then further assuming that some of the minor determinant of the order $M$ is
  non-zero, it results in $\bm{v}=0$. 
  This is contradiction and $\text{rank}\, \Psi <M$.
  Since the eigenvector $\bm{v}$ is non-vanishing, 
  we can assume without losing generality,
  say, $(\bm{v})_M=c\ne 0$ where 
  $\bm{v}=\begin{pmatrix} \bm{v}_R \\ c \\ \end{pmatrix}$.
  Then the vector $\bm{\psi}_M$ is expanded by the other MOs and
  \begin{align}
    \Psi &=\bar \Psi R,
   \notag \\
    \bar\Psi&=(\psi_1,\cdots,\psi_{M-1}))\in M(N,M-1),
    \notag \\
    R &= (I_{M-1},-\bm{v}_R/c)
    \in M(M-1,M).
  \end{align}
  It implies that the one-particle Hamiltonian is
  expressed by the reduced set of MOs,
  $|{\cal M}-1)=(|1),\cdots,|M-1))=| {\cal A} \rangle \bar \Psi $, as
  \begin{align} 
    {\cal H}_1 &= |{\cal M}-1)\bar h ( {\cal M} -1 |,
     \end{align}
     with
     \begin{align}
    \bar h &= R  h R ^{-1},
  \end{align}
  where
  $ |{\cal M)} = |{\cal A} \rangle \bar \Psi R= |{\cal M}-1 ) R$.

  Repeating the process 
  (and writing $\bar{M} \rightarrow M$ for simplicity),
  we generically assume
  \begin{align}
    \text{rank}\, \Psi &= M\le N,
    \ \
    \Psi \in M(N,M),
    \end{align}
    and 
    \begin{align}
    {\cal O} = \Psi ^\dagger \Psi &\in M(M,M), 
    \text{ is positive definite}.
  \end{align}
  It implies $\det {\cal O} \ne 0$ and all eigenvalues are positive.

  If $N>M$, this Hamiltonian ${\cal H} $ has $N-M$ degenerate zero modes
  since
  the zero energy condition,
  $( {\cal M} | \varphi \rangle =\Psi ^\dagger \bm{\varphi}=0$ for $|\varphi \rangle =|{\cal A} \rangle \bm{\varphi}$,
  gives $M$ linear conditions for $N$-dimensional vector $\bm{\varphi}$.

  It also obeys from a simple algebra for the secular equation [Eq.~(\ref{eq:det_reduce})]
  $\det_N( \lambda I_N-H)=\det_N( \lambda I_N-\Psi h\Psi ^\dagger )=
  \lambda ^{N-M}\det_M( \lambda I_M- h {\cal O}  )$.
  When the system is periodic,
  by considering 
  ${\cal H}_1$
  as a Hamiltonian of the energy bands,
  these zero modes give
  the flat band at zero energy, which is $N-M$ fold degenerated.

 \subsection{Hidden chiral symmetry \label{app1_2}}
Let us assume the MOs are 
classified into two classes,
  $A$ and $B$, as
  \begin{align}
    |{\cal M}) &=
    |{\cal M}_A) \oplus
    |{\cal M}_B)=|{\cal A} \rangle \Psi=|{\cal A} \rangle (\Psi_A,\Psi_B),
    \notag \\
    |{\cal M}_A )&= (|1)_A,\cdots,|M_A)_A=|{\cal A}\rangle \Psi_A,
    \notag\\
    |{\cal M}_B )&= (|1)_B,\cdots,|M_B)_B=|{\cal A}\rangle \Psi_B,
   \notag \\
    \Psi_A &\in M(N,M_A),
   \notag  \\
    \Psi_B &\in M(N,M_B),
  \end{align}
  where $M=M_A+M_B$.

  Further, assume each MO is \textit{non-overlapping} within the class and
  normalized as
  \begin{align}
    (i_A|j_A) &= \delta _{i_Aj_A},\ \Psi_A ^\dagger \Psi_A=I_{M_A},
    \\
    (i_B|j_B) &= \delta _{i_Bj_B},\ \Psi_B ^\dagger \Psi_B=I_{M_B}.
  \end{align}
  Now the overlap matrix $(\cal{M}|\cal{M})\in M(M,M)$  is written as
  \begin{align}
    {\cal O} &=  ({\cal M}|{\cal M}) =\mmat{I_{M_A}}{ ({\cal M}_A|{\cal M}_B) }{({\cal M}_B|{\cal M}_A)} {I_{M_B} }
    \notag \\
    &= \Psi ^\dagger \Psi
    =\mmat{I_{A}}{\Psi_A ^\dagger \Psi_B}{\Psi_B ^\dagger \Psi_{A}} {I_{M_B}}.
  \end{align}
  We then define a \textit{chiral symmetric} Hamiltonian for $M$ site system as
  \begin{align}
    h_C(\Psi_A,\Psi_B) &= {\cal O} -I_M=\Psi ^\dagger \Psi-I_M.
    \notag \\
    &= \mmat{O}{D ^\dagger}{D}{O},
  \notag  \\
    D &= \Psi _B ^\dagger \Psi _A\in M(M_B,M_A).
  \end{align}
  This is chiral symmetric by the chiral operator $\gamma $ as
  \begin{align}
    \acmt{h_C}{\gamma} &= h_C \gamma + \gamma h_C=0,
     \notag \\
    \gamma &= {\rm diag}\, (I_{M_A},-I_{M_B}).
    \end{align}
   Note that 
   $\gamma$ satisfies
 \begin{align} 
    {\rm Tr}\, \gamma = M_A-M_B, \hspace{1pt}
    \gamma ^2 = I_{M}.
  \end{align}
 
  As for the zero modes of $h_C$, which are also the zero modes of $h_C^2$,
  and 
  vice versa is true.
  Since $h_C^2$ commutes with $\gamma $, 
  they are labeled by the chirality as
  \begin{align} 
    h_C \Phi_\pm = 0, \hspace{1pt} \gamma \Phi_\pm &= \pm \Phi_\pm,
  \end{align}
  where
  \begin{align}
    \Phi_+ &= \mvec{\bm{w}_1}{0} , \Phi_- = \mvec{0}{\bm{w}_2},
  \end{align} 
   and
\begin{align} 
  D ^\dagger \bm{w}_1 &=0,\ D \bm{w}_2 =0.
\end{align} 

It implies
\begin{align}
  N_+^\gamma &= M_A-{\rm rank}\, D ^\dagger,
  \\
  N_-^\gamma &= M_B-{\rm rank}\,  D,
\end{align}
where $N_\pm ^\gamma $ is a number of the zero modes
for the standard chiral symmetry.
Since ${\rm rank}\,D=
{\rm rank}\,D ^\dagger $, it implies
\begin{align}
  N_+^\gamma -N_-^\gamma &= M_A-M_B.
\end{align}

\section{Relation to square-root Hamiltonian \label{app:sqr}}
Here we describe the relation between the MO representation 
and the square-root construction of the Hamiltonian~\cite{Arkinstall2017,Kremer2020,Mizoguchi2020_sq,Mizoguchi2021,Matsumoto2023,Mizoguchi2023}. 
In fact, using $\Psi$ and $\Psi^\dagger$ of Sec.~\ref{sec:mo},
one can construct the $(M+N)\times (M+N)$ Hamiltonian, 
\begin{align}
  H_{\rm sq} &= \mmat{O_{MM}}{\Psi ^\dagger }{\Psi  }{O_{NN}}\in M(N+M,N+M),
\end{align}
that satisfies
\begin{align}
  H_{\rm sq}^2 &=  \mmat{\Psi ^\dagger  \Psi   }{O_{MN}}{O_{NM}}{\Psi \Psi ^\dagger  }
  = {\rm diag}\, ({\cal O} ,H).
\end{align}
This square root $H_{\rm sq }$ is chiral symmetric
in a different sense from the chiral symmetry described by $\gamma$ and $\Gamma$.
To be specific, it satisfies
\begin{align}
  \acmt{\gamma _{sq}}{H_{\rm sq}}&= 0, \notag \\
  \gamma _{\rm sq} &= {\rm diag}\, (I_M,-I_N).
\end{align} 
The duality between the non-zero energy sector is written due to
the eigenvalue equation,
\begin{align}
  H_{\rm sq }\mvec{U}{V}&= 
  \mvec{U}{V} {\cal E}  ^{+1/2},
\end{align}
or equivalently, 
\begin{align}
  \mvec{U}{V}&=  
  H_{\rm sq }\mvec{U}{V} {\cal E}  ^{-1/2}.
\end{align}
The zero energy "flat bands" is given by the eigenstate of 
the  chiral operator $\gamma _{\rm sq }$
\begin{align}
  H_{\rm sq} \mvec{O_{MM}}{\varphi} &=0, \notag \\ 
  \gamma _{\rm sq} \mvec{O_{MM}}{\varphi} &= (-1)              
   \mvec{O_{MM}}{\varphi},
 \notag \\
  \varphi &= (\bm{\varphi}_1,\cdots,\bm{\varphi}_{Z}).
\end{align}

\begin{figure}[tb]
\begin{center}
\includegraphics[clip,width = 0.9\linewidth]{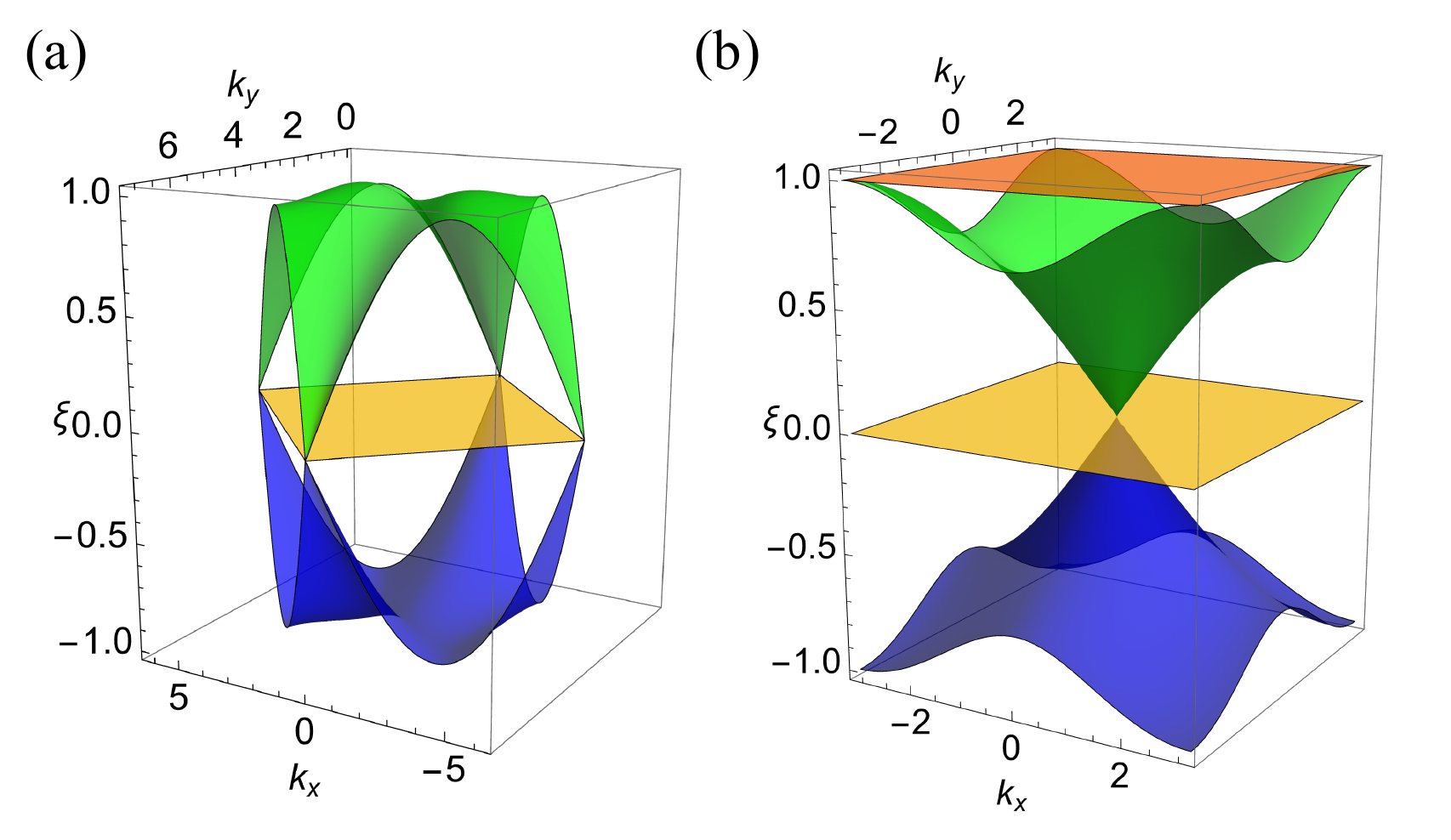}
\vspace{-10pt}
\caption{
The band structure of $\Gamma$ 
for (a) the kagome lattice
with 
$(\alpha_1,\alpha_2,\alpha_3,\beta_1,\beta_2, \beta_3)=\left(\frac{1}{\sqrt{3}},\frac{1}{\sqrt{3}},\frac{1}{\sqrt{3}},\frac{1}{\sqrt{3}},\frac{1}{\sqrt{3}},\frac{1}{\sqrt{3}}  \right)$
and (b) the square-ocgaton lattice with
$(\alpha_1,\alpha_2,\alpha_3,\alpha_4,\beta_1,\beta_2, \beta_3,\beta_4)=
\left(\frac{1}{\sqrt{2}},\frac{1}{\sqrt{2}},\frac{1}{\sqrt{2}},\frac{1}{\sqrt{2}},
\frac{1}{2},\frac{1}{2},\frac{1}{2},\frac{1}{2}
\right)$.
}
\label{fig:gamma_spac}
\end{center}
\vspace{-10pt}
\end{figure}

\section{Remark on the eigenvalues of $\Gamma$ \label{app:remark}}
In this appendix, 
we remark on the eigenvalues of 
$\Gamma$.
As we have seen in Sec.~\ref{sec:p-h},
the anti-commutator $\{\Gamma, H_C \} = 0$ 
and the Pythagoras relation $H_C^2 + \Gamma^2 = I_N$ 
require the particle-hole-symmetric eigenvalues of $H_C$.
Then, it is clear that the same constraint is assigned 
for the eigenvalues of $\Gamma$ by regarding $H_C$
as a chiral operator. 
To make the argument concrete, let us repeat the argument of Sec.~\ref{sec:p-h} with interchanging the role of $H_C$ and $\Gamma$.
Let $\bar{\Phi}_{\xi}$ be the eigenstate of $\Gamma$
satisfying $\Gamma\bar{\Phi}_{\xi}=\xi\bar{\Phi}_{\xi}$ ($\bar{\Phi}^\dagger_{\xi}\bar{\Phi}_{\xi} = 1$).
Then one can define a state $\bar{\Phi}^{C}_{\xi}:= H_C \bar{\Phi}_{\xi}$,
which satisfies $\Gamma \bar{\Phi}^{C}_{\xi} = \bar{\Phi}^{C}_{\xi} (-\xi)$.
Note that $(\bar{\Phi}^C_{\xi})^\dagger \bar{\Phi}^{C}_{\xi}=
\bar{\Phi}^{\dagger}_{\xi} (H_C)^2\bar{\Phi}_{\xi} 
= 1-\xi^2$.
Then, again, there are three possibilities:
\begin{enumerate}
  \item[(1)$^\prime$] 
  $\bar{\Phi}^{C}_{\xi} =0$ (its norm is zero). $\xi=\pm 1$.
  \item[(2)$^\prime$] 
  $\xi\ne\pm 1,0$, 
  $\bar{\Phi}^C_{\xi} $ is a non-vanishing eigenstate of 
  the energy $-\xi$. 
  The states $\Phi_{\xi}$ and $\Phi_{-\xi}$ are paired.
  \item[(3)$^\prime$] $\xi=0$, 
  $\Gamma \bar{\Phi}_{\xi=0}=0$. 
  Due to the Pythagoras relation, 
    $H_C^2 \bar{\Phi}_{\xi=0}= \bar{\Phi}_{\xi=0}$.
\end{enumerate}
This means that the eigenvalues of the 
$\Gamma$ with $\xi \neq 0, \pm 1$ appears in a pairwise manner.

This relation leads to an interesting consequence for the kagome lattice model,
particularly when $\alpha$s' and $\beta$s' are $1/\sqrt{3}$ [Fig.~\ref{fig:kagome}(b)].
In this case, $\Gamma$ of Eq.~(\ref{eq:gammaprime})
corresponds to 
the Hamiltonian for the breathing kagome lattice with hoppings on the upward triangles having the opposite sign to those on the downward triangles.
This ``staggered-sign" breathing kagome model 
has the particle-hole symmetric band structure~\cite{Essafi2017,Ezawa2018_kagome}.
In Fig.~\ref{fig:gamma_spac}(a), 
we show the band structure for $\Gamma$,
where we indeed see the positive and negative energy bands appear in a pairwise manner, 
and that the flat band is at zero energy,
which coincides with the (1$^\prime$)-(3$^\prime$). 
This fact is well-known, but, to our knowledge, its origin has not been clarified since the chiral symmetry due to the lattice structure is absent.
Note that $\xi = \pm 1$ also appears in a pairwise manner in this case. 
The Pythagoras relation indicates that they appear at K and K$^\prime$ where the Dirac points appear for $H$. 
Conversely, for $\xi$, 
the triple band touching point with linear dispersion around it, which is also called the spin-1 Dirac cone, occurs at 
$\bm{k} = 0$. At that point, $\varepsilon = 0,2$ (i.e., $E = \pm 1$) which also coincides with the Pythagoras relation.
It is also to be noted that the quadratic band touching between the flat band and the dispersive band occurs at this point. 

In Fig.~\ref{fig:gamma_spac}(b), 
we plot the band structure for $\Gamma$ for the square-octagon lattice model.
We see the spectrum is particle-hole symmetric, 
except for the flat band at $\xi = 1$. 
This is the consequence of (1)$^\prime$,
where the mode with $\xi = 1$ may not have its negative energy counterpart. 
In fact, in the present model, 
the additional flat band is enforced by the fact that Tr$ \Gamma =M_A -M_B = 1$ (at each $\bm{k}$).
Comparing Fig.~\ref{fig:gamma_spac}(b) 
with Fig.~\ref{fig:sqo}(b),
we find that the spin-1 Dirac cone for $\varepsilon$ appears 
at $\bm{k} = (\pi, \pi)$, 
where $\xi  = 0, \pm 1$.
Conversely, for $\xi$, 
the spin-1 Dirac cone, occurs at 
$\bm{k} = 0$, where $\varepsilon = 0, 1, 2$ (i.e., $E = -1,0,1$) and the quadratic band touching between the flat band and the dispersive band occurs. 

\bibliographystyle{apsrev4-2}
\bibliography{MO_chiral}
\end{document}

%% file: mydef.tex
\newcommand{\acmt}[2]{{\{}#1,#2{\}}}
\newcommand{\cmt}[2]{{[}#1,#2{]}}

\newcommand{\eas}[0]{\begin{eqnarray*}}
\newcommand{\eae}[0]{\end{eqnarray*}}
\newcommand{\les}[0]{\begin{equation}}
\newcommand{\lee}[0]{\end{equation}}
\newcommand{\leas}[0]{\begin{eqnarray}}
\newcommand{\leae}[0]{\end{eqnarray}}

\newcommand{\mmat}[4]
{
\left(
\begin{array}{cc}
#1 & #2 \\
#3 & #4 
\end{array}
\right)
}

\newcommand{\mvec}[2]
{
\left(
\begin{array}{c}
#1  \\
#2  
\end{array}
\right)
}

\newcommand{\mvecthree}[3]
{
  \left(
    \begin{array}{c}
#1  \\
#2  \\
#3  
    \end{array}
    \right)
}